\documentclass[twocolumn,prc,aps,amsmath,amssymb]{revtex4-2}
\usepackage{graphicx}
\usepackage{bm}
\usepackage{color}
\usepackage{float}
\usepackage{xcolor}
\usepackage{booktabs} 
\usepackage{multirow} 
\usepackage{makecell}

\begin{document}

\preprint{APS/123-QED}

\title{Search for the Efimov state near the 3$\alpha$ threshold}

\author{A. Baishya$^{1,2}$}
\email{abaishya@barc.gov.in}
\author{S. Santra$^{1,2}$}
\email{ssantra@barc.gov.in}
\author{T. Singh$^{1,2}$}
\author{P. C. Rout$^{1,2}$}
\author{A. Pal$^{1,2}$}
\author{H. Kumawat$^{1,2}$}
\author{T. Santhosh$^{3}$}
\author{P. Taya$^{1,2}$}
\author{M. Meher$^{1,2}$}
\author{Jyotisankar Das$^{4}$}

\affiliation{$^1$Nuclear Physics Division, Bhabha Atomic Research Centre, Mumbai 400085, India}
\affiliation{$^2$Homi Bhabha National Institute, Mumbai 400094, India}
\affiliation{$^3$Department of Nuclear and Atomic Physics, Tata Institute of Fundamental Research, Mumbai 400005, India}
\affiliation{$^4$Department of Physics, Sri Sathya Sai Institute of Higher Learning, Prasanthi Nilayam 515134, India}

\begin{abstract}
A high-precision experiment in search of the predicted Efimov state in $^{12}\mathrm{C}$ at 7.458 MeV excitation energy was performed. Using a state-of-the-art detector system and novel analysis techniques, it was possible to observe the Efimov state at the predicted energy level  above the 3$\alpha$ threshold with much better sensitivity in the $^{12} \mathrm{C}$ excitation energy spectrum compared to the existing data. The mutual $^{8} \mathrm{Be}$ resonance (91.84 keV) condition filters out a total of 21 probable Efimov state events around 7.458 MeV. With 2$\sigma$ confidence, it gives an upper limit of 0.014$\%$ for the Efimov state $\alpha$-decay width relative to that of the Hoyle state, which is about an order of magnitude smaller than the latest upper limit found in the literature. This observation was supported by a new penetrability calculation assuming a
relatively extended structure of the Efimov state compared to the Hoyle state. The effect of the Efimov state was also explored in a nuclear astrophysical scenario, where the triple-$\alpha$ reaction rate, including both the Hoyle state and the Efimov state, was found to be larger than the allowed limit, while the temperature dependence of the combined rate was found to be compatible with the helium shell flash criterion of the AGB stars.
\end{abstract}

\maketitle

The Efimov effect represents one of the most striking manifestations of quantum mechanics in three-body systems. Originally proposed in nuclear physics, the effect occurs when a system of three particles exhibits an infinite series of bound states under specific conditions, despite the two-body subsystem being unbound. This phenomenon, first described by Vitaly Efimov~\cite{Efimov1971}, hinges on the delicate interplay of long-range three-body forces, which emerge when the two-body systems have a large s-wave scattering length. While the Efimov effect has been confirmed experimentally in ultracold~\cite{Kraemer2006} atomic systems, its presence in nuclear systems remains an intriguing yet unverified possibility. 

In nuclear physics, the $^{12}\mathrm{C}$ nucleus serves as a compelling candidate for investigating Efimov states due to its prominent $\alpha$-cluster structure. Of particular interest is the Hoyle state, an excited 0$^+$ state at 7.65 MeV, which plays a crucial role in the triple-$\alpha$ process, facilitating the synthesis of carbon in stars. The Hoyle state is widely considered a paradigmatic example of an $\alpha$-cluster structure, with theoretical models depicting it as a resonance of three weakly interacting $^4\mathrm{He}$ nuclei. In his original paper~\cite{Efimov1971}, Efimov initially suggested the Hoyle state to be an Efimov state candidate. But the Hoyle state fails to satisfy the necessary conditions for the Efimov effect to occur; for example, Suno {\it et al.}~\cite{Suno2015} showed the three-body Efimov potential to be too weak compared to the Coulomb and nuclear potential. However, whether the Hoyle state or any nearby resonance can be classified as an Efimov state remains a subject of debate.

\begin{figure}[h]
\begin{center}
\includegraphics[width=8.3cm,height=5.6cm]{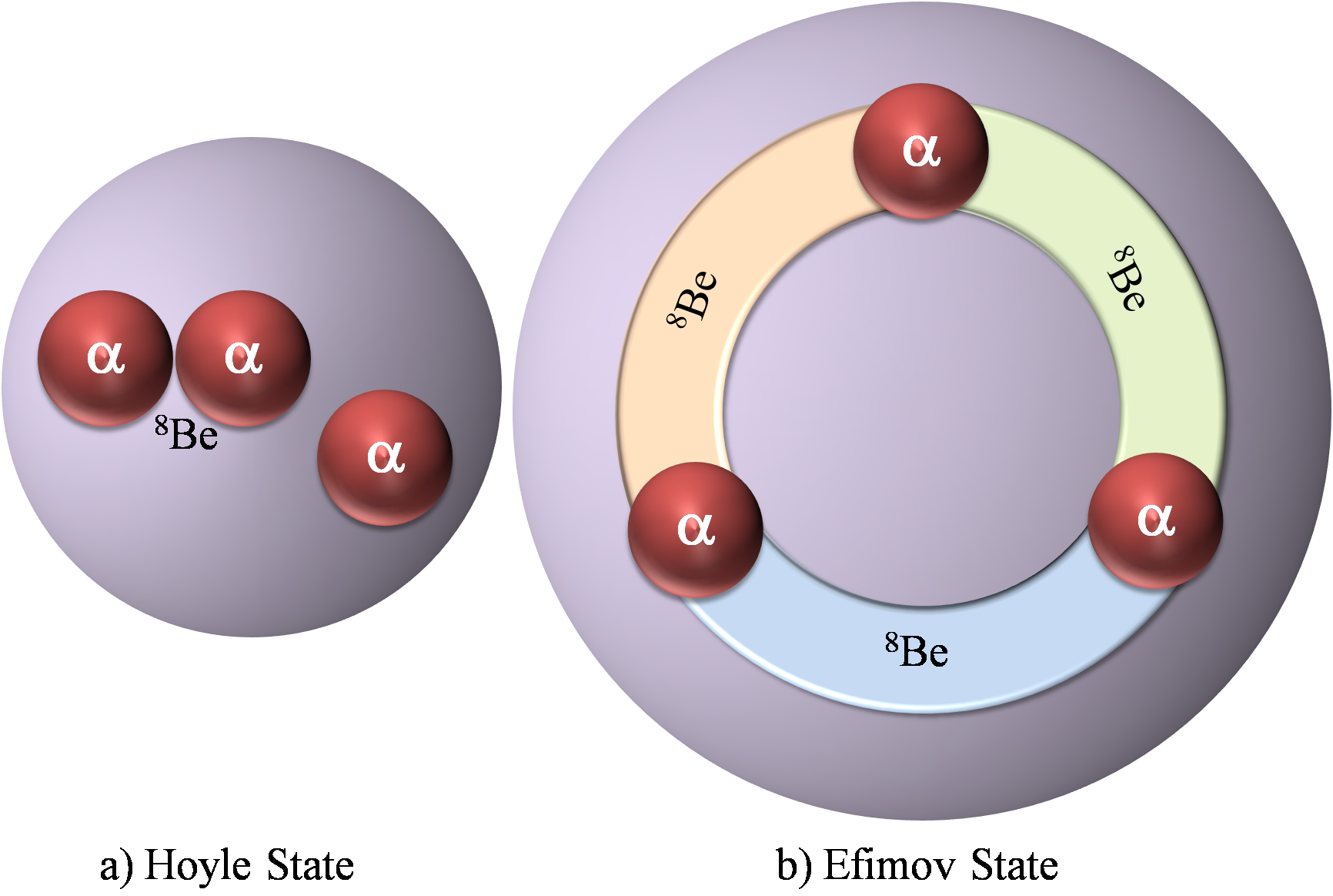}
\end{center}
\caption{\label{fig:diagram} Artistic representation of a) the Hoyle state depicted by the bent-arm structure~\cite{Epelbaum2012}  and b) the Efimov state with an extended structure and any two $\alpha$ particles  having mutual $^8\mathrm{Be}$ resonance. } 
\end{figure}

Recently, theoretical predictions have identified a potential Efimov-like state at an excitation energy of approximately 7.458 MeV in $^{12}\mathrm{C}$~\cite{Zheng2018,Zheng2020}, from kinematical arguments. They considered the Efimov effect to appear in the $^{12}\mathrm{C}$ nucleus due to mutual $\alpha$ exchange between the other two remaining $\alpha$ nuclei in the nucleus (Fig.~\ref{fig:diagram}); as a result, all three $\alpha$ particles are in mutual $^8\mathrm{Be}$ resonance. From this simple kinematic consideration, the excitation energy of such a state can also be easily realized.

\begin{equation}
\label{eq:Ex}
E_{Efimov} = \frac{2}{3}\sum_{i\neq j}^{3} E_{ij} + E_{th} 
\end{equation}

Putting $E_{ij} = 0.092$ MeV and $E_{th} = 7.274$ MeV, the 3$\alpha$ breakup thresholds, together, one arrives at $E_{Efimov} = 7.458$ MeV. Such a state, if observed, could provide a new window into the physics of three-body interactions and the interplay of nuclear forces near the threshold. However, theoretical challenges, such as disentangling the Efimov effect from standard nuclear and Coulomb interactions, as well as experimental limitations in accessing these states, have complicated its verification.

There have been a few experimental searches for the probable Efimov state at the aforementioned energy of 7.458 MeV using a variety of techniques. Zhang {\it et al.}~\cite{Zhang2019} using heavy ion collisions at very high energies (35 MeV/A), claimed an upper limit for the ratio of Efimov state $\alpha$-decay to Hoyle state $\alpha$-decay in $^{12}\mathrm{C}$ to be 0.3$\%$. Although they employed an interesting analysis technique by introducing the mixing method and representation of the decay probability by a fit function, the inherent multi-channel nature of the reactions involved means the existence of random $\alpha$ contribution is unavoidable. In another work, $\beta$-delayed charged-particle spectroscopy combined with $\gamma$-spectroscopy and multi-particle coincidence methods have been employed to probe near-threshold Efimov state in $^{12}\mathrm{C}$ by Bishop {\it et al.}~\cite{Bishop2021}. They claimed no possibility of an Efimov state at 7.458 MeV. The strict constraint was put in place with the help of $\gamma$-decay measurement of $\beta$-decay of $^{12}\mathrm{N}$ and $^{12}\mathrm{B}$. However, only the $\alpha$-decay measurement provides an upper limit for the Efimov state $\alpha$-decay compared to the Hoyle state $\alpha$-decay of 0.69$\%$. Cardella {\it et al.}~\cite{Cardella2022} with the CHIMERA detector set-up, have performed measurements of coincident three-$\alpha$ events, while simulations have provided crucial insights into the decay modes of hypothetical Efimov states. Their results are consistent with no Efimov state $\alpha$-decay in sequential decay (SD) mode, but, for the direct decay (DD) mode from the Efimov state, an upper limit of 0.2$\%$ was suggested. However, due to poor resolution in the $^{12}\mathrm{C}$ excitation energy spectrum for both of the above studies, clear separation of 7.654 MeV Hoyle state and 7.458 MeV Efimov state is not visible in either case. It could be worthwhile to perform an experiment with better energy and angular resolution and higher statistics.

The search for an Efimov state in $^{12}\mathrm{C}$ is not merely a quest to confirm a theoretical prediction but also a gateway to deeper insights of the nature of nuclear forces and how they manifest into alpha clustering in different nuclei. This work aims to advance the exploration of the Efimov phenomenon in $^{12}\mathrm{C}$ by employing cutting-edge experimental techniques and robust theoretical frameworks. The results presented herein have implications not only for nuclear structure but also for our understanding of nucleosynthesis in astrophysical environments. \\

\begin{figure}[h]
\begin{center}
\includegraphics[width=8.5cm,height=6cm]{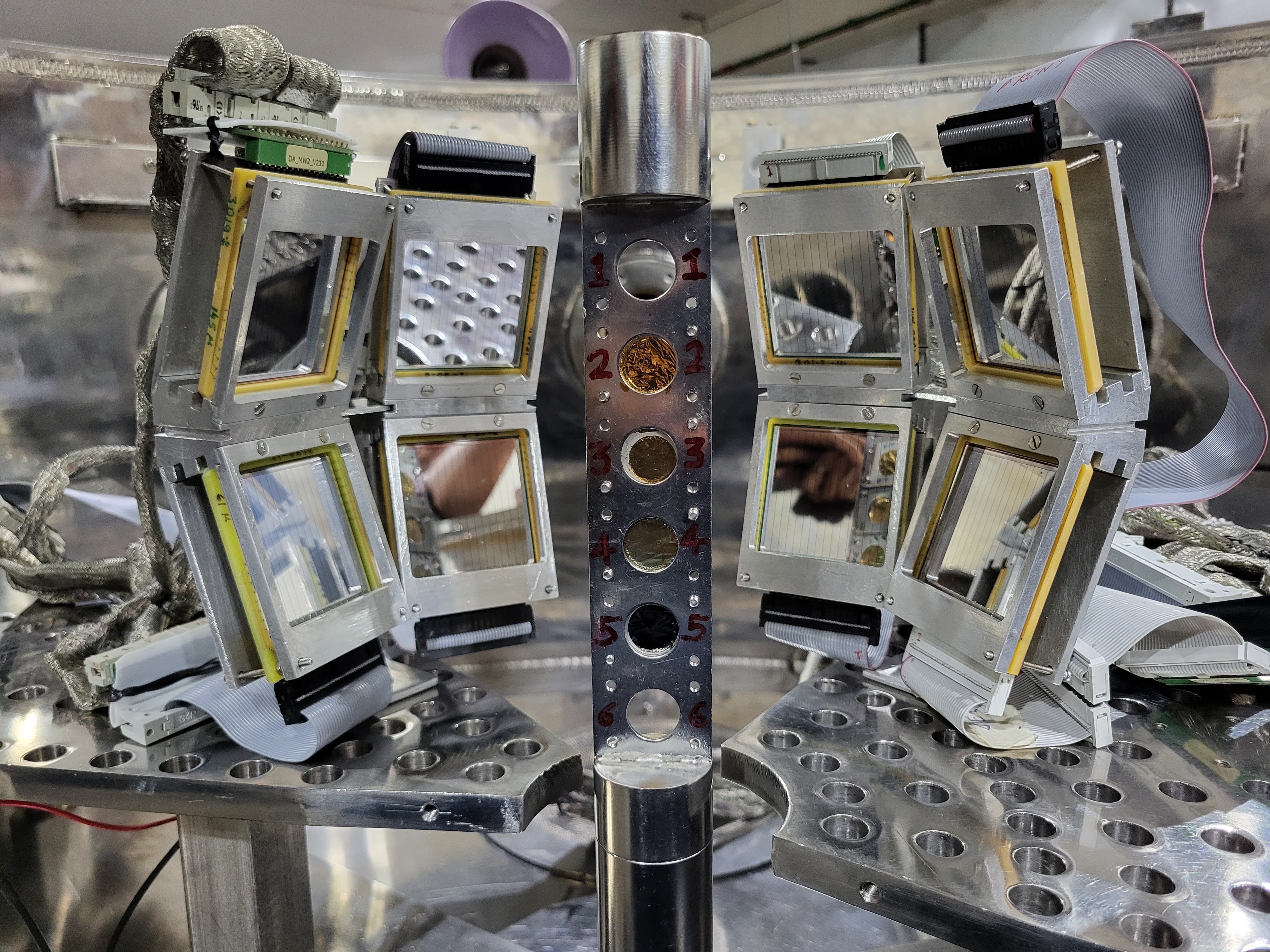}
\end{center}
\caption{\label{fig:setup} Experimental setup consisting of eight DSSDs arranged in two arrays and placed symmetrically with respect to the beam axis.} 
\end{figure}

The experiment was conducted at the 14UD BARC-TIFR Pelletron Accelerator Facility, Mumbai. A pulsed $^{12}\mathrm{C}$ beam of energy of 57 MeV was incident on a self-supporting natural carbon target of thickness $\approx25~\mu \text{g/cm}^2$. The choice of $^{12}\mathrm{C}$ as both the projectile and target made the reaction symmetric in the center-of-mass frame. As the desired reaction can populate the resonances of $^{12}\mathrm{C}$ above the $\alpha$ threshold, three $\alpha$ nuclei from the excited $^{12}\mathrm{C}$ are expected. For detecting these breakup $\alpha$ particles, charged particle detectors were employed. The detector setup consisted of eight double-sided silicon strip detectors (DSSD) divided into two arrays, each consisting of four DSSDs and placed symmetrically at $50^\circ$ with respect to the beam axis (Fig.~\ref{fig:setup}). Each DSSD has 16 vertical front strips and 16 horizontal back strips, dividing the active detection area into 256 smaller pixels that act as individual detectors. This allows one to measure both the energy and the position of the detected charged particles. The DSSD mounts were designed in such a way that their centers are spherically symmetric with respect to the target center. Each DSSD center was 123 mm from the target center. Simulation was carried out beforehand to place the arrays in their optimum position. The trigger for event acquisition was generated such that whenever both the arrays fire simultaneously, an event is recorded. The event rate was kept below 1 kHz to reduce contribution from random coincidences. Each DSSD was calibrated for energy measurement using the known energy $\alpha$-particles emitted from the Am-Pu and the Th $\alpha$ sources for lower energy points along with elastic counts from the $^{12}\mathrm{C} \left(^{197}\mathrm{Au},^{197}\mathrm{Au} \right) ^{12}\mathrm{C}$ reaction at 35 MeV for higher energy points.

\begin{figure}[h]
\begin{center}
\includegraphics[trim=0.1cm 0.3cm 0.55cm 0.3cm, clip,scale=0.49]{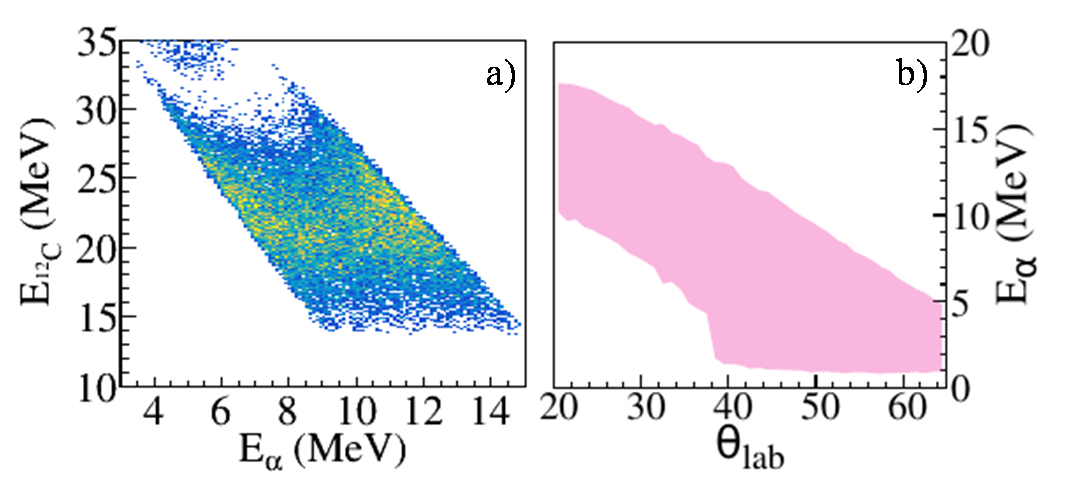}
\end{center}
\caption{\label{fig:kingate1} a) Simulated kinematic gate that represents the correlation between the energy of the breakup $\alpha$ particles and the recoil $^{12}\mathrm{C}$. b) Simulated kinematic gate that gives the allowed range of the breakup $\alpha$ particle energies as a function of their detected lab angle.} 
\end{figure}

The novelty of the present detection technique lies in the use of single E-detectors in place of $\Delta$E+E telescopes in order to reduce the event rate and accommodate more E-detectors to increase the solid angle coverage. However, particle identification was ensured by employing several kinematic conditions while analyzing the data. This methodology was first verified by analyzing the data acquired from a short experiment using telescopes. The data analysis steps are as follows. The collected events were first filtered out by demanding a total of four hits in coincidence from the two arrays, with three from one of the arrays and the remaining one from the other array. A 2D kinematic gate was generated using a Monte-Carlo three-body breakup simulation code, which correlates a genuine breakup $\alpha$ particle energy to the recoil $^{12}\mathrm{C}$ energy, as shown in Fig.~\ref{fig:kingate1}(a). Another kinematic gate generated by the simulation code correlates the detected $\alpha$ lab angle to the allowed  energy range of that breakup $\alpha$ particle from the desired reaction, as shown in Fig.~\ref{fig:kingate1}(b). Further, timing gate, momentum conservation gate, and energy conservation gate (both for 7.654 MeV+g.s. and 7.654 MeV+4.443 MeV channels) were applied. Events satisfying the above conditions were considered truly kinematically complete and were chosen for building the excitation energy spectrum. Using a DSSD-type charged particle detector allowed us to measure not only the energy of the particles but their hit positions as well. From the energy and position information (spherical $\theta$ and $\phi$ angles), one can reconstruct the velocity vectors of the detected charged particles simply as,

\begin{multline}
\label{eq:valpha}
\overrightarrow{v_{\alpha_i}} = \sqrt{\frac{2E_{\alpha_i}}{m_{\alpha_i}}} (sin\theta_{\alpha_i}cos\phi_{\alpha_i}\hat{i} + sin\theta_{\alpha_i}sin\phi_{\alpha_i}\hat{j} \\ + cos\theta_{\alpha_i}\hat{k})
\end{multline}

 Once the velocity vectors are reconstructed, the excitation energy can be calculated by using the Eq.~\ref{eq:Ex}, where $E_{ij}$ is the relative energy between the $i^{th}$ and $j^{th}$ $\alpha$ particles and is given by,

\begin{equation}
\label{eq:Eij}
E_{ij} = \frac{1}{2}\mu_{\alpha \alpha} \left( \overrightarrow{v_{\alpha_i}} - \overrightarrow{v_{\alpha_j}} \right)^2
\end{equation}

where $\mu_{\alpha \alpha}$ is the reduced mass for the $\alpha+\alpha$ system, $ \overrightarrow{v_{\alpha_i}}$ and  $\overrightarrow{v_{\alpha_j}}$ the velocity of the $i^{th}$ and $j^{th}$ $\alpha$ particles.

\begin{figure}[tbh!]
\begin{center}
\includegraphics[trim=0.2cm 0.3cm 1.7cm 0.4cm,clip,width=8.6cm]{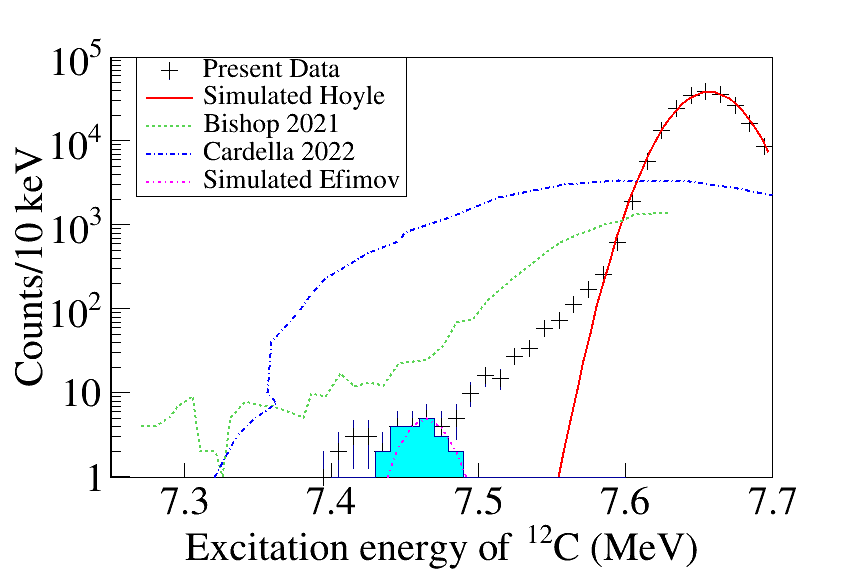}
\end{center}
\caption{\label{fig:Ex} Excitation energy spectrum of $^{12}\mathrm{C}$ generated by complete kinematic 3$\alpha$ events. The cyan-filled area represents potential Efimov events. } 
\end{figure}

The distribution of $^{12}\mathrm{C}$ excitation energy, extracted from the present data, is shown as symbols in Fig.~\ref{fig:Ex}, and is compared with those from Refs.~\cite{Bishop2021,Cardella2022} along with the simulated spectra for the Hoyle state and the Efimov state generated by a Monte-Carlo breakup code utilizing the experimental details such as detector distance, angle, energy resolution, etc. Due to much better energy and angular resolution of the used detector system, the Hoyle state peak at 7.65 MeV is by far the sharpest and is most certainly an advantage in these kinds of rare studies. 

To identify the potential Efimov state counts, one can use two methods as used in Ref.~\cite{Bishop2021} and \cite{Cardella2022} respectively. First, the method of Ref.~\cite{Cardella2022} which is same as Ref.~\cite{Zhang2019} was used, where events are filtered with the condition that all three $\alpha$-pair relative energies correspond to 92 keV $^8$Be resonance. In the present study, the breakup-simulation code provided the relative energy range to implement the gate, which is 60 keV to 120 keV. In this way, a total of 21 potential Efimov state events were found, which is shown under the cyan-colored shaded area in Fig.~\ref{fig:Ex}. Applying the Poisson statistics, the 2$\sigma$ upper limit for the Efimov state counts comes out to be $\approx$31. Now, the $\Gamma_{\alpha}$ of the Efimov state, with respect to the Hoyle state, can be determined from the potential Efimov counts as follows:

\begin{equation}
\label{eq:upperlimit1}
\frac{\Gamma_{\alpha}^{ES}}{\Gamma_{\alpha}^{HS}} < \frac{P_{HS}}{P_{ES}} \times \frac{N_{ES}}{N_{HS}}
\end{equation}

where $N_{ES}$ and $N_{HS}$ are the experimentally observed potential counts for the Efimov state and the Hoyle state, respectively, and $P_{ES}$ and $P_{HS}$ are the population probabilities of the Efimov state and the Hoyle state, respectively. The population probability of a particular excited state can be approximated by the Boltzmann factor ($e^{-\frac{E_i}{k_BT}}$). For the present case, the ratio $\frac{P_{HS}}{P_{ES}}$ is approximately 1. Thus, 

\begin{equation}
\label{eq:upperlimit}
\frac{\Gamma_{\alpha}^{ES}}{\Gamma_{\alpha}^{HS}} \lessapprox  \frac{N_{ES}}{N_{HS}}
\end{equation}

With almost 2.21$\times$10$^5$ Hoyle state events, this gives the upper limit of $\alpha$-decay width ratio of the Efimov state to the Hoyle state ($\frac{\Gamma_{\alpha}^{ES}}{\Gamma_{\alpha}^{HS}}$) to be $\approx$0.014$\%$. This upper limit is a huge improvement over the limits of 0.69$\%$ and 0.2$\%$ put forward by Ref.~\cite{Bishop2021} and Ref.~\cite{Cardella2022}, respectively. 

Now, if the first method~\cite{Bishop2021} is used to identify potential Efimov state events, which is to consider the counts above the simulated Gaussian curve but with energies $<$7.5 MeV, one arrives at a number of 43. The 2$\sigma$ upper limit of $\frac{\Gamma_{\alpha}^{ES}}{\Gamma_{\alpha}^{HS}}$ thus becomes $\approx$0.025$\%$, still a big improvement over the previous studies. This improvement is mostly due to the improved resolution of the present data. The counts in the region of potential Efimov state energy in both Refs.~\cite{Bishop2021} and \cite{Cardella2022} are much higher compared to the present data, and these counts are well explained by their respective Gaussian fits, meaning due to poorer resolution, the tail of the Hoyle state extends to much lower energy compared to the present study.

One interesting way to look at the above results is in terms of $\alpha$ penetrability of the Efimov state compared to the Hoyle state. The authors have previously used this technique to elucidate different configurations of $\alpha$ structure for the Hoyle state~\cite{Abhi2021} and it was concluded that from purely penetrability point of view, equilateral triangular structure is the most probable one. For the present penetrability calculation, the nuclear interaction used in Ref. \cite{Abhi2021} was retained. The penetrability ratio of Efimov state to Hoyle state was found to be of the same order ($\sim10^{-8}$) as that shown in Ref.~\cite{Zheng2018}, which is much smaller than, but still consistent with, our experimentally obtained upper limit of 0.014$\%$. 

\begin{figure}[tbh!]
\begin{center}
\includegraphics[scale=1.05]{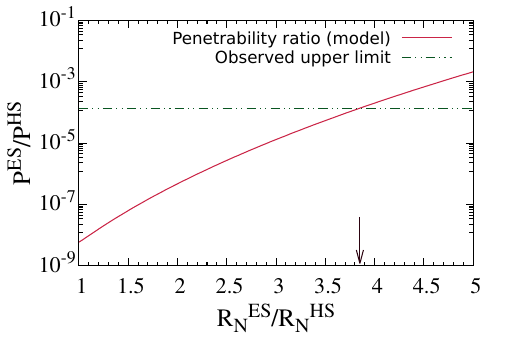}
\end{center}
\caption{\label{fig:penetrability} Ratio of `penetrability of the Efimov state' to `that of the Hoyle state' as a function of the ratio of `the inner turning radius of Efimov state' to `that of the Hoyle state', $\frac{R_N^{ES}}{R_N^{HS}}$.} 
\end{figure}

It is well accepted in the literature that Efimov state are spatially diffused and extended structures. So, in principle, for calculating the penetrability of an $\alpha$ particle from an Efimov state, one can increase the inner turning radius, $R_N$ (where, the kinetic energy of the tunneling $\alpha$ equals to the combined potential due to Coulomb and nuclear interaction between the $\alpha$ and the residual $^8$Be) to get a higher value of penetrability that matches the observed experimental upper limit. Fig.~\ref{fig:penetrability} shows the ratio of penetrability of the Efimov state to the Hoyle state  as a function of the ratio of the inner turning radius of the Efimov state to that of the Hoyle state, $\frac{R_N^{ES}}{R_N^{HS}}$. Interestingly, it is observed that, for $\frac{R_N^{ES}}{R_N^{HS}} = 3.85$, the penetrability ratio matches our experimental upper limit ($0.014\%$).  Thus, assuming the Efimov state to be relatively spatially extended compared to the Hoyle state, one can explain the present experimental upper limit. This allows one to place an indirect upper bound on the spatial size of the Efimov state.

Finally, the potential impact of the Efimov state on the triple-$\alpha$ reaction rate by calculating the triple-$\alpha$ reaction rate using the formalism given in Refs.~\cite{Angulo1999,Tsumura2021} was investigated. Details of the formalism along with the used parameters are given in Appendix~\ref{app:rate}. A reaction rate code was developed in modern Fortran, with numerical integration carried out using the {\sc QUADPACK} library~\cite{quadpack}, and the Coulomb functions calculated using the {\sc COUL90} subroutine~\cite{coul90}. 

\begin{figure}[tbh!]
\begin{center}
\includegraphics[trim={0.15cm 0.5cm 0.2cm 0.0cm},clip,scale=1.04]{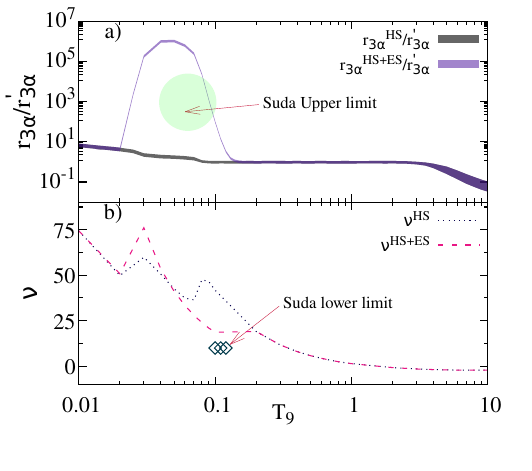}
\end{center}
\caption{\label{fig:r3a} (a) Ratio of the triple-$\alpha$ reaction rate calculated for the Hoyle state and Hoyle state plus the Efimov state to the rate by Tsumura {\it et al.} ($r^{\prime}_{3\alpha}$). The Suda upper limit for the same ratio is also shown in a green circle, with its size roughly representing the uncertainty of the limit value. (b) Temperature dependence of the triple-$\alpha$ rate, $\nu$ $= \frac{d\, \mathrm{ln}(r_{3\alpha})}{d\, \mathrm{ln}(T)}$, is shown along with the Suda lower limit for the same.} 
\end{figure}

Fig.~\ref{fig:r3a}(a) shows the ratio of the triple-$\alpha$ reaction rate ($r_{3\alpha}$) calculated for the Hoyle state and the Efimov state to the rate ($r^{\prime}_{3\alpha}$) calculated by Tsumura {\it et al.}~\cite{Tsumura2021}, which, in addition to the Hoyle state, considers two additional resonances, 9.64 MeV $3_1^-$ resonance and 9.87 MeV $2_2^+$ resonance, with updated decay parameters. Three different cases are considered: 1) the reaction rate only through the Hoyle state (solid line), 2) the combined reaction rate through both the Hoyle state and the Efimov state (dash-dot-dot line), and 3) a constraining point on the ratio given by Suda {\it et al.}~\cite{Suda2011} (green circle) by considering the reaction rate ($< 10^{-29} \quad \text{cm}^6.\text{mol}^{-2}.\text{s}^{-1}$ at $10^{7.8}$ K or $T_9 = 0.063$) required by a star to enter the red giant phase. The associated uncertainties in the first two cases were calculated using a Monte-Carlo uncertainty propagation code. Triple-$\alpha$ reaction rates through other resonances (the 9.64 MeV $3_1^-$ resonance and the 9.87 MeV $2_2^+$ resonance) are not included, as they are negligible for $T_9 < 5$ compared to the rate through the Hoyle state (Fig.~\ref{fig:r3a}a). The Suda limit was found to be lower by two orders of magnitude than the calculated combined rate. This indicates that the $\alpha$-decay width may be even lower than the upper limit observed from the present measurement. However, the Suda limit itself may be relaxed under various conditions, such as enhanced CNO cycling, increased core mass at ignition, elevated conductive opacities, additional mixing mechanisms (e.g., rotational or magnetic), and reduced metallicity.

In addition to the triple-$\alpha$ rate,  the temperature dependence of the triple-$\alpha$ rate, defined as $\nu=\frac{d\, \mathrm{ln}(r_{3\alpha})}{d\, \mathrm{ln}(T)}$, through the Hoyle state plus the Efimov state resonances, is  shown in Fig.~\ref{fig:r3a}(b). A $\nu$ value of at least 10 at $T \sim (1–1.2) \times 10^8$ K is required for the stars to undergo helium shell flashes, and in turn, the third dredge-up to happen, which is essential for producing carbon and s-process elements (observed in AGB stars)~\cite{Suda2011}. Clearly, the temperature dependence of the combined Hoyle state plus the Efimov state rate is higher than the Suda lower limit. This indicates that the present calculated rate satisfies the stars having helium shell flashes even though the absolute rate is higher at low temperature. Currently, the Suda limit is given only at one temperature ($T_9 = 0.063$), but with a spectrum of limiting points, it will help achieve better constraint on the triple-$\alpha$ reaction rate.

In summary, a precision experiment was performed to search for the Efimov state near the $3\alpha$ threshold in $^{12}\mathrm{C}$. The state-of-the-art experiment was conducted by employing large arrays of DSSD charged particle detectors in combination with an advanced data acquisition setup. Applying novel analysis techniques, an upper limit of 0.014\% for the $\alpha$-decay width ratio of the Efimov state to the Hoyle state was achieved, which is currently the most stringent limit, more than an order of magnitude lower than the existing limit~\cite{Cardella2022}. 
The $\alpha$-penetrability calculation showed that an extended structure of $^{12}$C with a radius approximately 3.85 times larger than that of the Hoyle state could account for the measured upper limit. 

The observed Efimov state significantly enhances the triple-$\alpha$ reaction rate. The total rate via the Hoyle state and the Efimov state, computed using a newly developed reaction rate code, exceeds the latest evaluation~\cite{Tsumura2021} by over six orders of magnitude, yet remains within two orders of the Suda limit~\cite{Suda2011}. This hints at a potentially lower $\alpha$-decay width for the Efimov state. Notably, the combined rate has a steeper temperature dependence than the Suda lower bound and still meets the helium shell flash criterion for AGB stars.

The present study, with the new upper bound on the Efimov state leading to refined constraints on the triple-$\alpha$ reaction rate, is an important step towards the understanding of the abundance of $^{12}$C, while unveiling a new horizon for simulating further studies of such an exotic state.

The authors are thankful to the BARC-TIFR Pelletron crew for the smooth operation of the accelerator during the long experiment.

\appendix
\section{\label{app:rate}Triple-$\alpha$ Reaction Rate}

The expression for the triple-$\alpha$ reaction rate~\cite{Angulo1999,Tsumura2021} is given as,

\begin{multline}
\label{eq:3alpha}
r_{3\alpha} = N_A^2 \langle \sigma v \rangle^{\alpha \alpha \alpha} = 3 N_A \left( \frac{8 \pi \hbar}{\mu_{\alpha \alpha}^2} \right) \left( \frac{\mu_{\alpha \alpha}}{2 \pi k_B T} \right)^{3/2} \times \\ \int_0^\infty \frac{\sigma_{\alpha \alpha} \left(E\right)}{\Gamma_\alpha (^{8} \mathrm{Be}, E)} \exp\left(-\frac{E}{k_B T}\right) N_A \langle \sigma v \rangle^{\alpha^{8} \mathrm{Be}} E \, dE.
\end{multline}

Here, $r_{3\alpha}$ is in $\text{cm}^6.\text{mol}^{-2}.\text{s}^{-1}$, $N_A$ is the Avogadro number, $\mu_{\alpha \alpha}$ is the reduced mass of the $\alpha+\alpha$ system, and $E$ is the energy above the $\alpha+\alpha$ threshold. The cross section of the $\alpha+\alpha$ scattering is given by,

\begin{equation}
\label{eq:sigalpalp}
\sigma_{\alpha \alpha}(E) = \frac{2\pi}{\kappa^2} \frac{\Gamma_\alpha(^8\text{Be}, E)^2}{(E - E_r^{8\text{Be}})^2 + \Gamma_\alpha(^8\text{Be}, E)^2/4},
\end{equation}

\begin{equation}
\label{eq:Gammaalp}
\Gamma_\alpha(^8\text{Be}, E) = \Gamma_\alpha(^8\text{Be}) \frac{P_0(E)}{P_0(E_r^{8\text{Be}})},
\end{equation}

where $\kappa$ is the relative wave number for the $\alpha+\alpha$ system, $\frac{\sqrt{2\mu_{\alpha \alpha}E}}{\hbar}$. $E^{^8\mathrm{BE}}_r = 91.84$ keV and $\Gamma_{\alpha}(^8 \mathrm{Be}) = 5.57$ eV are the energy and width of the $^8 \mathrm{Be}$ ground state, respectively. $P_l(E)$ being the penetration factor associated with the relative angular momentum $l$ at the channel radius of $a = 1.4(A_1^{1/3}+A_2^{1/3}) = 4.4\, \mathrm{fm}$, is given by $\frac{1}{F_l^2+G_l^2}$, $F_l$ and $G_l$ are the regular and irregular Coulomb functions, respectively.

If a bound $^8 \mathrm{Be}$ is formed at an energy E different from the resonance energy, $E^{^8\mathrm{BE}}_r$, the reaction rate, $N_A \langle \sigma v \rangle^{\alpha^{8} \mathrm{Be}}$, for the second step of fusing the third $\alpha$ and the $^8\mathrm{Be}$ is expressed as,

\begin{multline}
\label{eq:NAsigmav}
N_A \langle \sigma v \rangle^{\alpha ^8\mathrm{Be}} = N_A \left( \frac{8\pi\hbar}{\mu_{\alpha ^8\mathrm{Be}}^2} \right) \left( \frac{\mu_{\alpha ^8\mathrm{Be}}}{2\pi k_B T} \right)^{3/2} \times \\  \int_0^\infty \sigma_{\alpha ^8\mathrm{Be}} \left( E', E \right) \exp\left(-\frac{E'}{k_B T} \right) E' \, dE', 
\end{multline}

where $\mu_{\alpha ^8\mathrm{Be}}$ is the reduced mass of the $\alpha + ^8\mathrm{Be}$ system, and $E'$ is the energy above the $\alpha + ^8\mathrm{Be}$ threshold. $\sigma_{\alpha ^8\mathrm{Be}} \left(E', E\right)$ is given by,

\begin{multline}
\label{eq:sigalpBe}
\sigma_{\alpha^{8}\mathrm{Be}}\left(E',E\right) = \left(2J + 1\right) \frac{\pi \hbar^{2}}{2 \mu_{\alpha^{8}\mathrm{Be}}} \times \\ \frac{\Gamma_{\alpha}(^{12}\mathrm{C}^{J}, E') \Gamma_{\gamma(E\lambda)}(^{12}\mathrm{C}^{J}, E' + E)}{\left(E' - E_{r}^{J} + E - E_{r}^{8}\mathrm{Be}\right)^{2} + \frac{1}{4} \Gamma(^{12}\mathrm{C}^{J}, E', E)^{2}},
\end{multline}

where $J$ is spin of a $3\alpha$ resonance in $^{12} \mathrm{C}$ at an energy of $E_r^J$. The $\alpha$-decay width is given by $\Gamma_{\alpha} \left( ^{12} \mathrm{C}^J, E' \right) = \Gamma_{\alpha} \left( ^{12} \mathrm{C}^J \right) \frac{P_J\left(E'\right)}{P_J \left(E_r^J \right)}$, while the $\gamma$-decay width via the $E\lambda$ transition is given by $\Gamma_{\gamma \left( E \lambda \right)} \left(^{12} \mathrm{C}^J, E' + E \right) = \Gamma_{\gamma \left( E \lambda \right)} \left(^{12} \mathrm{C}^J \right) \frac{\left( E_T + E' + E - E_r^{^8 \mathrm{Be}} \right)^{2\lambda+1}}{ \left( E_T + E_r^J \right)^{2\lambda+1}}$, and the total width $\Gamma = \Gamma_{\alpha} + \Gamma_{\gamma}$. The $\gamma$ threshold energy is defined as $E_T = 7.366 - E_x$ MeV, where $E_x$ is the excitation energy of the final state of the $E\lambda$ decay, which for the present case is the $2_1^+$ level at 4.443 MeV; thus, $E_T = 2.92$ MeV. $E_r$, $\Gamma_{\alpha} \left( ^{12} \mathrm{C} \right)$, and $\Gamma_{\gamma \left( E \lambda \right)} \left(^{12} \mathrm{C}^J \right)$ are given in Table.~\ref{tab:nuclear_params}. 

\begin{table}[h]
\centering
\caption{$^{12} \mathrm{C}$ resonance parameters used to calculate the triple alpha reaction rate in the present work.}
\label{tab:nuclear_params}
\small
\begin{tabular}{ccccc}
\toprule
\textbf{Nucleus} & \textbf{$J_{n}^{\pi}$} & \textbf{$E_{r}$ (keV)} & \textbf{$\Gamma_{\alpha}$ (eV)} & \textbf{$\Gamma_{\gamma(E\lambda)}$ (meV)} \\
\midrule
$^{8} \mathrm{Be}$ 
    & $0^{+}_{1}$ 
    & \makecell{91.84 \\ $\pm$ 0.04} 
    & \makecell{5.57 \\ $\pm$ 0.25} 
    & -- \\
\midrule
\multirow{2}{*}{$^{12} \mathrm{C}$} 
    & $0^{+}_{2}$ 
    & \makecell{287.7 \\ $\pm$ 0.2} 
    & \makecell{9.3 \\ $\pm$ 0.9} 
    & \makecell{3.81 \\ $\pm$ 0.39} \\
\cmidrule(lr){2-5}
    & $0^{+} \left( \text{Efimov} \right)$ 
    & \makecell{91.84 \\ $\pm$ 0.04} 
    & \makecell{$8.84 \times 10^{-4}$ \\ $\pm 2.11 \times 10^{-4}$} 
    & $5.28 \times 10^{-8}$ \\
\bottomrule
\end{tabular}
\end{table}

For the ES, the resonance parameters are not known experimentally; hence, they were calculated as follows: i) $\Gamma_{\alpha}^{ES}$ was obtained using Eq.~\ref{eq:upperlimit}, thus, $\Gamma_{\alpha}^{ES} = \Gamma_{\alpha}^{HS} \times \frac{21}{2.21\times10^5} = 9.3 (eV)\times 9.5\times10^{-5} = 8.84 \times 10^{-4}$ eV, ii) $\Gamma_{\gamma}^{ES}$ was calculated using the method given by Depastas {\it et al.}~\cite{Depastas2025},  $\Gamma_{\gamma}^{ES} = \Gamma_{\gamma}^{HS} \left( \frac{E_T + E_r^{ES}}{E_T + E_r^{HS}} \right)^5 \frac{\Gamma_{\alpha} \left( ^{12} \mathrm{C}^{0_2^+}, E_r^{ES} \right)}{\Gamma_{\alpha} \left( ^{12} \mathrm{C}^{0_2^+}, E_r^{HS} \right)} = 3.81 (meV)\times \left( \frac{3.014}{3.21} \right)^5 \times 1.9 \times 10^{-8} = 5.28 \times 10^{-8}$ meV.

\end{document}